# Vapor-cell frequency reference for short-wavelength transitions in neutral calcium

JENNIFER TAYLOR,[1] BRYAN HEMINGWAY,[1] JAMES HANSSEN,[1] THOMAS B. SWANSON,[1] STEVEN PEIL[1,*]

[1]Time Service Department, United States Naval Observatory—3450 Massachusetts Avenue N.W., Washington, D.C. 20392
*Corresponding author: steven.peil@navy.mil

We have characterized the molecular tellurium (Te$_2$) spectrum in the vicinity of the 423nm $^1S_0$-$^1P_1$ and the 431nm $^3P_1$-$^3P_0$ transitions in neutral calcium. These transitions are relevant to optical clocks for atomic-beam characterization and cooling (423nm) and enhanced detection (431nm). The use of a Te$_2$ vapor cell as a frequency reference has many advantages over other laser stabilization techniques, and we discuss an application to measuring the instability due to the second-order Doppler shift in a calcium beam clock.

## 1. INTRODUCTION

**Operational Optical Clocks**

Optical atomic clocks have resulted in dramatically improved performance over their microwave predecessors since their introduction in the past decade [1]. While state-of-the-art ion-trap and optical-lattice clocks provide the highest accuracy [2-4], there have been quite a few efforts in recent years to develop more robust, even portable, optical clock technology. These "operational" optical clocks could address applications that do not necessarily require absolute frequencies but that could benefit from greater availability. Examples include operational clocks used for generation of precise time [5], space clocks for applications such as global navigation [6], and field clocks for relativistic geodesy [7].

While there has been some effort to engineer optical-lattice systems [8, 9], in many cases the approach to building robust optical clocks has been to carry out spectroscopy on an atomic beam [10-12]. This is because laser technology is arguably the most serious obstacle to robust, continuous operation for an atomic clock (or any other atomic sensor). Aside from a frequency comb, a thermal (uncooled) beam clock can be constructed with a single laser for interrogation of the clock transition, although a second laser to enhance detection can dramatically improve the performance. In comparison, an optical-lattice system requires the clock laser, a laser for creating the lattice, two lasers for cooling and up to two lasers for repumping.

*Atomic Beam Optical Clock*

Optical clocks use alkaline-earth atoms or other elements with the same $^1S_0$ ground-state electronic structure, but only a subset of these elements are practical for atomic-beam clocks. The available time to excite the clock transition in a beam is short compared to the interaction time available in a trapped-atom system. For saturated absorption spectroscopy, measurement-time limited resonances on order of 500kHz are typical [10,11], while Ramsey-Bordé interrogation can generate linewidths from 50kHz to below 5kHz [11,12]. Therefore, the extremely narrow (typically 1Hz or less), doubly forbidden $^1S_0$-$^3P_0$ transitions cannot be fully utilized in a beam clock; in fact, these weak transitions are difficult to excite in the short interaction time available, typically requiring on order of 1W of laser power. The singly forbidden $^1S_0$-$^3P_1$ transitions are therefore better suited. In many $^1S_0$ ground-state atoms, the singly forbidden $^1S_0$-$^3P_1$ transition is relatively broad, and only a few elements have transitions with spectral widths that are fairly well matched with the interaction times one can achieve in a beam. These include strontium, calcium, and magnesium with $^1S_0$-$^3P_1$ transitions that have natural widths of 7kHz, 375Hz and 40Hz respectively.

*Calcium*

Of these, the 657nm $^1S_0$-$^3P_1$ transition in calcium ($^{40}$Ca) has some beneficial features. The 375Hz width is narrow enough not to limit the resolution for even the best Ramsey-Bordé measurements, and the laser wavelength required for the strong $^1S_0$-$^1P_1$ transition used for cooling and other applications is much more convenient than for magnesium. In fact, calcium would be especially well suited for a cooled-beam clock, where resolutions approaching the natural width of the transition are achievable with the additional complexity of only one laser. Calcium was the subject of early precision optical spectroscopy [13,14] and subsequent cooled-atom optical clock efforts [15,16], but it was determined not to be ideally suited as a primary frequency standard. Yet, for the reasons presented, it has been the subject of recent efforts to build robust optical frequency standards that emphasize stability over accuracy.

One challenge with optical clocks based on calcium, or other $^1S_0$ ground-state atoms, is the availability of frequency references for stabilizing lasers tuned to the necessary optical transitions. While the interrogation laser is locked to the clock transition, lasers for other transitions need some other reference. In optical-clock atoms, there is typically a strong $^1S_0$-$^1P_1$ transition with a wavelength in the blue part of the spectrum. In calcium, this transition occurs at 422.792nm, with a natural width of about 35MHz; an energy level diagram is shown in Figure 1. This line can be used for cooling [15,16], detection [11], or characterization of the oven and beam [17]. The $^3P_1$–$^3P_0$ transition in calcium at 431nm shown in the figure is even better to use for detection; it can be driven to force an atom excited by the clock laser to scatter many blue photons, while the majority of atoms remaining in the ground state do not contribute a background signal.

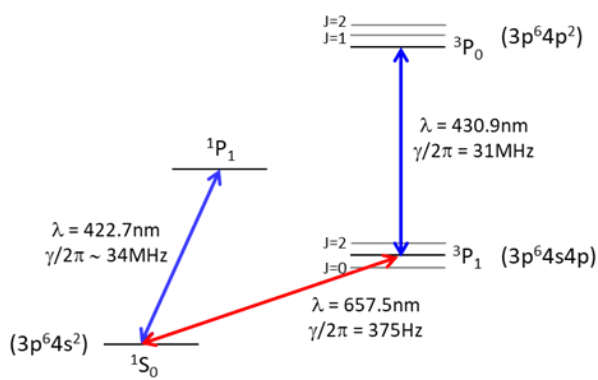

Figure 1. (Color online.) Partial energy level diagram for calcium ($^{40}$Ca), showing the clock transition at 657nm and the blue transitions at 423nm and 431nm. Atoms in the $^3P_1$ excited clock state can be driven to a higher energy $^3P_0$ state with 431nm light. Atoms in that state can only decay back to $^3P_1$ due to angular momentum selection rules, making the 431nm transition a cycling transition useful for enhancing the detection signal.

*Laser Frequency Stabilization*

Simple vapor cells are ideal for stabilizing the frequencies of lasers that are subsequently used for spectroscopy, cooling, or trapping, especially for applications that require long-term frequency stability or immunity to vibrations. They are routinely used for alkali-atom systems where the cells contain a vapor of the same element that is being used in the main apparatus. However, vapor cells are not an option for calcium or other $^1S_0$ ground-state atoms; at the necessary temperatures (of order 500°C) to create a reasonable vapor pressure for spectroscopy, calcium interacts with most glasses, rendering the glass opaque and the cell useless. Stabilizing these laser frequencies for calcium has typically been done using a Fabry-Perot cavity or the atomic beam itself, but there are drawbacks to both of these techniques. A good cavity requires a high-vacuum chamber and thorough vibration isolation, especially if it is to be used during transport, and it will exhibit frequency drift at some level over long times. Locking to the atomic beam has the drawback of introducing additional vacuum windows and laser light to the clock vacuum chamber. Also, there are additional challenges to using the atomic beam to stabilize a laser to be used for the 431nm detection line, since that signal relies on two resonant excitations. Less common frequency references that have been utilized for these wavelengths include heat pipes [18], vacuum vessels in which the spectroscopy windows are positioned away from a high vapor region and held at a lower temperature, and hollow-cathode lamps [19]. Heat pipes are somewhat elaborate, requiring fabrication of a custom vacuum chamber, and hollow-cathode lamps can suffer from poorly understood frequency shifts and are known to be susceptible to plasma instabilities [20].

## 2. TELLURIUM VAPOR CELL REFERENCE

### A. Introduction

Molecular vapor cells have an abundance of transitions available to serve as frequency references because of the many rotational and vibrational modes. While molecular iodine has been a workhorse for frequency calibration and stabilization at wavelengths over much of the visible spectrum, its dissociation limit near 500nm precludes its use at blue wavelengths. Like iodine the tellurium dimer (Te$_2$) has many transitions due to molecular excitations, but tellurium can serve as a frequency reference well into the violet [21]. A sufficient vapor pressure for spectroscopy can be generated by heating tellurium to around 500°C, and it can be done using a standard glass cell since there are no deleterious reactions as in the case of calcium. Tellurium has been investigated generally [22] and used as a frequency reference for work on ytterbium [23] and barium ions [24] and for repumping in neutral strontium [25].

*Experimental Setup*

We investigate the tellurium spectrum using (saturated) absorption and modulation transfer spectroscopy (MTS) in a heated vapor cell while simultaneously driving the calcium transitions of interest. Light from a Ti:sapphire laser is passed through a commercial frequency doubling cavity with a BBO (barium borate) crystal. For 2W of near-IR light, we generate as much as 0.5W of blue light. The frequency tuning of the Ti:sapphire laser required to reach both of the calcium transitions of interest is accomplished using the birefringent filter. The doubling cavity has a modest bandwidth of about 1nm for a fixed crystal orientation; re-orienting the BBO crystal allows the doubler to be tuned between the two wavelengths of interest, 8nm apart, though with reduced efficiency. The generated blue light is fiber coupled, sent through a fiber splitter, and sent to both a tellurium spectroscopy setup and a calcium atomic beam, as shown in Figure 2.

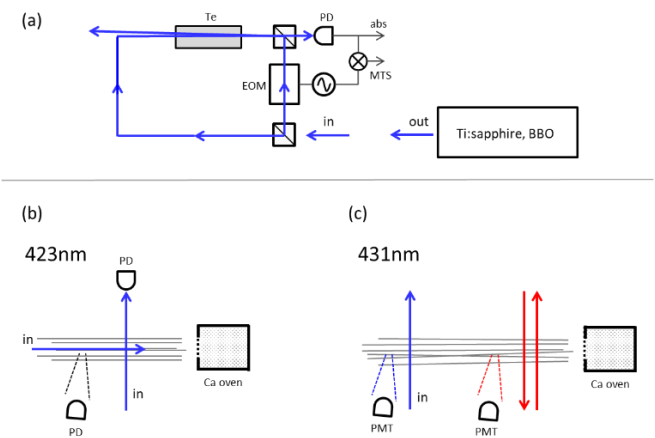

Figure 2. (Color online.) Illustration of setup used to measure the Te$_2$ spectrum in the vicinity of the corresponding calcium transitions. (a) Illustration of absorption and modulation transfer spectroscopy setup to measure Te$_2$ spectrum. Bottom: Arrangement of laser beams used to measure 423nm resonance (b) and 431nm resonance (c) in a calcium beam. The laser transverse to the atomic beam in (b) can have known frequency shifts introduced for calibration. Thick (blue/red) lines with arrow heads represent laser light; thin black lines in (a) represent electronic signals; PD – photodiode; PMT – photomultiplier tube; abs – absorption signal output; MTS – modulation transfer spectroscopy output.

Two different arrangements are used to look at the calcium response. For the 423nm $^1S_0$-$^1P_1$ transition, shown in Figure 2(b), we send several mW of laser light transverse to the atomic beam and measure the transmitted power. This gives an absorption resonance at the true calcium frequency, without any Doppler shift. The laser light in this path can have known frequency shifts introduced using an acousto-optic modulator for calibration. Simultaneously, we send several mW of light counter-propagating to the atomic beam and collect fluorescence from the atoms. This resonance is shifted due to the average velocity of the beam and is broadened by the distribution of velocities. Observing the 431nm $^3P_1$–$^3P_0$ transition requires prior excitation of the 657nm clock transition. In that case, excitation of the 657nm transition is implemented in a spectroscopy region of the vacuum chamber and further downstream we apply the blue light and collect fluorescence. This is illustrated in Figure 2(c).

## B. Characterization of Te$_2$ Spectrum at 423nm

The spectrum of tellurium in the vicinity of the 423nm transition in calcium is shown in Figure 3. The three plots on the left show the absorption of laser light by tellurium (top), the demodulated MTS signals (middle), and the calcium transverse absorption and longitudinal fluorescence resonances (bottom). There are three tellurium lines in the ~5GHz-wide spectrum shown. In the right half of the figure, higher resolution curves of a saturated absorption dip and the corresponding MTS signal for the strongest tellurium line are shown.

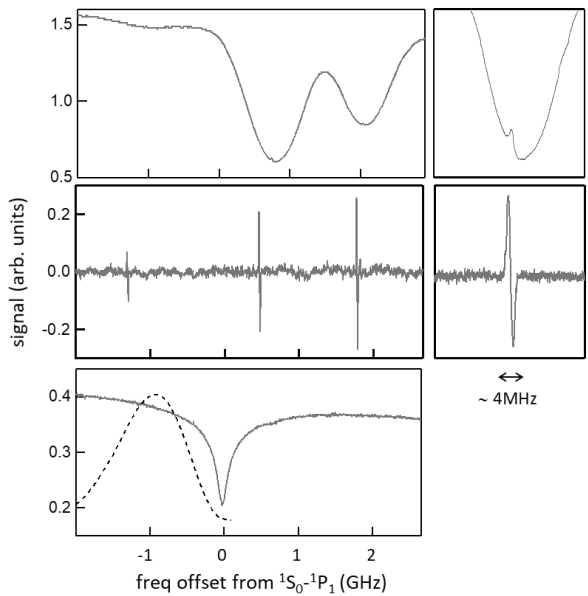

Figure 3. Tellurium and calcium resonances near 422.79nm. Left: Three tellurium lines in the vicinity of the calcium $^1S_0$-$^1P_1$ transition are shown via absorption curves (top) and demodulated MTS signals (middle). The calcium resonances for absorption of laser light transverse to the atomic beam direction (solid line) and fluorescence from light propagating against the atoms' trajectory (dashed line) are shown in the bottom plot. The fluorescence curve is scaled to fit on the plot. The slope on the background of the absorption curve reflects a slight frequency dependence in the doubler output. Barely visible are saturated absorption dips in the absorption resonances in the top plot and a signal from calcium-44 in the shoulder of the calcium absorption curve. Right: A closer look at some of the features in the plots on the left by scanning over a smaller range of frequencies.

For saturation of the tellurium resonance at this wavelength, a ~1mm beam waist with ~10mW of power distributed between the pump and probe is required for the MTS setup. For the traces in the figure, the 10cm Te$_2$ cell was heated to 400°C. The transition closest to the 423nm line is relatively strong, and if the cell were heated much higher, the optically thick vapor would prevent the pump and probe beams from interacting with the same atoms and the saturation signal would be lost. On the other hand, the Te$_2$ signal in the plot that is lower in frequency corresponds to a weaker transition, and this could easily be enhanced by heating the cell further.

As shown in Figure 3, the closest Te$_2$ resonance to the calcium 423nm transition is a strong line about 500MHz higher in frequency. This line can be used as a frequency reference, but for some purposes a different transition may be preferable. While the weaker Te$_2$ signal at the lower frequency in the plot is further from the $^1S_0$-$^1P_1$ resonance, it is closer to the Doppler-shifted frequency for conventional thermal atomic-beam velocities. This Te$_2$ line may be preferable for frequency locking for laser cooling or for measurements where a reference for fast atoms is desired.

### Application to Measuring Instability of 2$^{nd}$-Order Doppler Shift

An example of this latter application is a measurement of the long-term stability of the second-order Doppler shift of a beam clock. The long-term performance of a thermal beam clock is expected to be limited by instabilities in Doppler shifts. Residual first-order Doppler shifts are likely to be a big obstacle to a stable long-term system, but the size and stability of these residual first-order effects are difficult to estimate. Even if first-order shifts could be completely eliminated in a thermal beam, the instability caused by variations in the second-order Doppler shift would still exist and could be a problem; with $\Delta\nu_{(2nd\ Dopp)} = v^2/2c^2 \sim 10^{-12}$ for a thermal calcium beam, the temperature of the atom source needs to be stable to about $10^{-5}$ (about 5mK) to avoid introducing frequency instability above $10^{-17}$.

What is achievable in terms of second-order Doppler stability in a thermal beam has not yet been investigated, but as the interest in using these systems as robust clocks grows a determination of the limits imposed by thermal instabilities would clearly be valuable. The stability floor imposed by variations in the second-order Doppler shift should be at a level that makes it nontrivial to measure directly, especially in the presence of other effects that could be comparable or larger. It should be possible to determine the stability of the atomic beam velocity by looking at the first-order Doppler shift, which can be maximized and made larger than any other effect by sending a laser beam counter-propagating to the atomic beam path. Measuring the stability of the first-order Doppler shift can then be leveraged to determine the stability of the second-order Doppler shift, where the lever arm is $c/v$, about $5\times10^5$ for a calcium beam.

Since Doppler shifts are due to the atomic velocity, the fractional frequency fluctuations are the same for all atomic transitions, and for calcium they can be measured with better signal-to-noise (S/N) using the 423nm transition than using the 657nm clock transition. By measuring the frequency difference between the 423nm Doppler-shifted resonance and the nearest tellurium line, the long-term stability of the beam velocity can be determined. With a width of ~1GHz, measurement of the Doppler shifted resonance with S/N~100 should allow a measurement of the instability of the second-order Doppler shift that integrates as $\sigma \sim 10^{-13}/\tau^{1/2}$, where $\sigma$ is the Allan deviation characterizing the degree of frequency fluctuations and $\tau$ is the averaging time. This could lead to interesting limits at measurement times of days. The tellurium line closest to the Doppler shifted calcium resonance provides an ideal reference for this proposed measurement, as the proximity in frequency and the long-term stability make it better suited than other alternatives.

### C. Tellurium Spectrum at 431nm

*Calcium detection resonance*

In addition to the strong 423nm transition, calcium has a blue line at 431nm that is particularly well suited for detection of atoms excited by the 657nm clock laser. The clock transition couples ground state $3p^64s^2$ $^1S_0$ atoms to the $3p^64s4p$ $^3P_1$ state, which has a long lifetime of order 1ms. For the fast atomic velocities in a thermal beam, most atoms in this excited sate will never emit a photon in the chamber, let alone scatter multiple photons. On average, detection of the clock transition by collecting 657nm photons results in a signal from less than 1 photon per atom.

The $3p^64p^2$ $^3P_0$ state that lies 431nm away from the excited clock state has a short lifetime of tens of ns (with corresponding natural width of order 30MHz) before decaying back to $3p^64s4p$ $^3P_1$. Because of angular momentum selection rules, this state can only decay back to the $^3P_1$ state via electric dipole coupling. This $^3P_0$-$^3P_1$ transition therefore constitutes a cycling transition that can scatter many blue photons for every atom excited by the 657nm interrogation laser. Furthermore, it only couples to atoms excited by the 657nm light; the detection signal from this transition is free from the background of the majority of atoms in the beam that are not excited with the clock laser. Of course, it is not free of laser-light background; design of collection optics to image the atoms in the path of the 431nm beam and to block light coming from other locations helps to minimize the stray laser light collected.

The advantage of using the 431nm transition for detection is illustrated in Figure 4, which shows two calcium resonances obtained with the two different detection methods. The curves correspond to single sweeps of the clock laser frequency over the entire Doppler profile and were acquired in 50ms through a 10kHz low-pass filter, with no additional averaging. The resonances on the left are shown on the same scale and are labeled by which photons are collected; the 657nm signal is rescaled in the trace on the right.

The data were taken simultaneously using similar parameters (photomultiplier tubes with similar sensitivities, bandpass filters with similar insertion loss at the corresponding wavelength). The improvement in signal to noise from using the 431nm transition of a factor of ~10 is clearly visible, and further optimization of the blue detection signal should be possible.

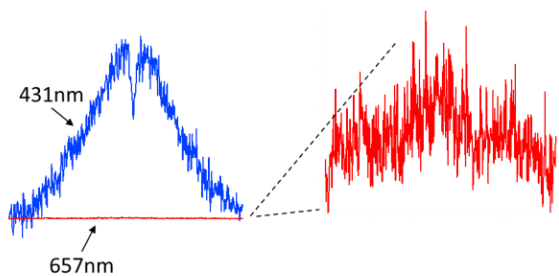

Figure 4. (Color online.) Left: Comparison of atomic resonance curves obtained by detecting emission at 657nm (red) and laser induced fluorescence at 431nm (blue) on the same scale. Right: Re-scaled 657nm signal. Each trace was acquired with a single 50ms scan of the 657nm laser frequency and with similar parameters for the photomultiplier tubes used. The difference in signal-to-noise ratio between the two methods is about a factor of 10.

*Tellurium resonances*

The distribution of tellurium resonances in the vicinity of the 431nm calcium transition is not as favorable as in the 423nm case, with the closest line being more than 1GHz away. This makes calibrating the frequency differences between the calcium and tellurium resonances using the modest frequency bandwidth of an AOM challenging, and instead we used a Fizeau-interferometer-based wavelength meter to measure the different resonances with uncertainties of order 100MHz.

The measured wavelength of the $^3P_0$-$^3P_1$ calcium resonance is 430.8951nm. This resonance and the nearest Te$_2$ lines are shown in Figure 5 (recall the setup used to make this measurement in Figure 2), with more than 10GHz of the nearby spectrum displayed. A Te$_2$ line is located at a frequency that is about 1.3GHz higher than the calcium transition, at a wavelength of 430.8943nm. This is a somewhat weak transition, requiring a vapor cell temperature above 550°C to achieve significant optical depth. A stronger Te$_2$ transition is located at a wavelength of 430.8970nm, a full 3GHz below the calcium detection line. In this case, strong absorption of the pump/probe beams occurs as low as 400°C.

The frequency separations between these tellurium lines and the calcium resonance are not very convenient. Additionally, saturation intensities for the tellurium resonances at this wavelength are higher than at 423nm, requiring on order of 30mW distributed between pump and probe in order to produce a saturation dip for the strong transition at 430.8970nm; even with 100mW of power, no saturation was observable in the weaker 430.8943nm transition. Locking to the strong resonance using a frequency sideband generated using a high-frequency (~3GHz) EOM would provide a stabilized laser for the calcium detection resonance.

### 3. SUMMARY

In conclusion, we have investigated using a tellurium vapor cell as a frequency reference for short-wavelength transitions in neutral calcium. A molecular transition measured in a vapor cell provides a long-term stable frequency reference and can be made compact and robust, to the point of being compatible with operation during transit. This is in contrast to most other frequency stabilization methods such as optical cavities, which exhibit long-term drift and sensitivity to vibrations. Tellurium offers a particularly convenient reference line for fast calcium atoms in a thermal beam. Disadvantages include the amount of optical power required to saturate the tellurium transition and, for systems where total electrical power is a concern, heating a cell to 500°C could be an obstacle. The required optical power is much less if a simple absorption signal can be used for frequency stabilization. Finally, we point out that a tellurium vapor cell should be useful for a frequency reference or laser stabilization for the $^1S_0$-$^1P_1$ transitions in strontium at 461nm and ytterbium at 399nm, as well as for the magic wavelength for a magnesium optical lattice at 468nm [26].

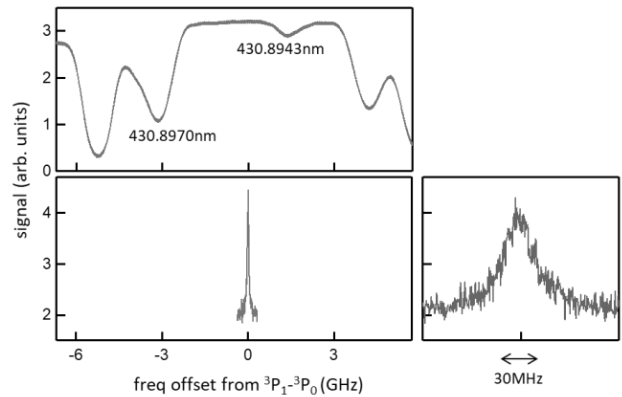

Figure 5. Tellurium spectrum in the vicinity of λ=430.8951nm. The calcium $^3P_0$-$^3P_1$ detection resonance (bottom – atomic response as blue

laser is scanned, with 657nm laser on resonance) and the nearest tellurium absorption signals (top). The tellurium vapor cell was heated to 470°C for this data. A higher resolution scan of the calcium resonance is shown on the right.

**Acknowledgments**. The benefits of using the 431nm transition in calcium for detection were pointed out to us by Chris Oates, and we benefited from discussions with others working on the calcium optical frequency reference at NIST, especially Judith Olson and Richard Fox. Clock Development at USNO receives funding from the Office of Naval Research.